\documentclass[12pt]{article}

\usepackage{sbc-template}
\usepackage{graphicx,url}
\usepackage[utf8]{inputenc}
\usepackage{authblk}
\usepackage{amsmath}
\usepackage{booktabs}
\usepackage[most]{tcolorbox}
\usepackage{amssymb}
\usepackage{makecell} 
\usepackage[breaklinks=true]{hyperref}
\usepackage{breakcites}
\usepackage{tablefootnote}
\usepackage{threeparttable}

\newcommand{\onedot}{\hspace{0.1em}.}
\def\eg{\emph{e.g}\onedot}

\def\etal{\emph{et al}\onedot}

\begin{document}
     
\sloppy

\title{Exploring the Role of Women in Hugging Face Organizations}

\author{Maria Tubella Salinas\inst{1}, Alexandra González\inst{1}, Silverio Martínez-Fernández\inst{1} }

\address{Universitat Politècnica de Catalunya
 \email{\{maria.tubella@estudiantat.upc.edu, alexandra.gonzalez.alvarez,silverio.martinez\}@upc.edu}
}

\maketitle

\begin{abstract}
\textbf{Background}: Despite its impact on innovation, gender diversity remains far from fully being achieved in open-source projects.
\textbf{Aims}: We examine gender diversity in Hugging Face (HF) organizations, investigating its impact on innovation and team dynamics in open-source development projects.
\textbf{Method}: We conducted a repository mining study, focusing on ML model development projects on HF, to explore the involvement of women in collaborative processes.
\textbf{Results}: Women are highly underrepresented in both organizations and commits distribution, which is also found when analyzing individual developers. 
\textbf{Conclusions}: Addressing gender disparities is essential to create more equitable, diverse, and inclusive open-source ecosystems. 
\end{abstract}


\section{Introduction}
Diversity is crucial in Software Engineering (SE), fostering innovation and effective problem-solving \cite{book}. More specifically, gender diversity is being explored as, in recent decades, the software development field has been marked by a significant gender imbalance, with men overwhelmingly representing the majority of the workforce. For instance, only 26.8\% of the people hired in Google's Tech area in 2023 were women \footnote{\href{https://static.googleusercontent.com/media/about.google/ca//belonging/diversity-annual-report/2023/static/pdfs/google_2023_diversity_annual_report.pdf?cachebust=2943cac}{https://static.googleusercontent.com/media/about.google/ca//belonging/diversity-annual-report/2023/ static/pdfs/google\_2023\_diversity\_annual\_report.pdf?cachebust=2943cac}}. This lack of gender diversity is prevalent, not only at the individual level but often across entire teams and organizations. In a survey conducted by \cite{ben-and-diff}, most respondents indicated that there is a lack of support for diversity in the SE field, with outdated patterns continuing to prevail, alongside instances of sexism and prejudice. Such gender homogeneity stands in contrast to the needs of software projects today, which rely heavily on collaboration, creativity, and diverse perspectives. Research shows that team diversity, including gender diversity, can enhance problem-solving capabilities, foster innovation, and lead to higher-quality outputs \cite{gendertenure}. Yet, many women who contribute meaningfully to the field remain underrepresented, often lacking visibility and recognition for their contributions.

Hugging Face (HF) is an advanced Machine Learning (ML) open-source platform known for its extensive library of pre-trained models, particularly in natural language processing and computer vision. Its ecosystem supports seamless model deployment, fine-tuning, and community-driven development, fostering rapid innovation. We focused our study in this platform due to its rapid growth and increasing popularity in both research and industry. Its versatility and active community make it an appealing choice for exploring modern ML workflows and tendencies.

On the HF platform, users can collaborate through entities called \textit{Organizations} \footnote{\href{https://huggingface.co/organizations}{https://huggingface.co/organizations}}. These HF Organizations serve as collaborative spaces where teams, research groups, companies, and other entities can jointly manage ML projects. By establishing an organization, members gain a shared environment for publishing, managing, and maintaining models, datasets, and other ML assets as a team. Any HF member can join one or more organizations, each of which can create and maintain an unlimited number of models. HF's openness is a foundational concept, essential for accurately interpreting the results.
 
Our research addresses multiple facets of gender composition and its potential influence on open-source ML projects. First, we aim to understand how HF organizations are structured in terms of gender diversity and whether this composition correlates with the popularity of the models they develop, measured by metrics such as downloads and likes. Secondly, we explore how women contribute within these organizations, particularly regarding the frequency and impact of their code commits. Analyzing commit patterns can offer insights into whether men or women tend to serve as primary contributors, potentially shaping project direction. Finally, we examine the nature of interactions between male and female members within these teams. By analyzing these dimensions, we aim to reveal how gender diversity manifests in HF organizations and explore its potential connections to organizational behavior and ML model success. This work ultimately aims to highlight the roles and contributions of women within this community, with broader implications for gender diversity in open-source software development. 

\section{Related Work}
Extensive research has been conducted to study diversity across work groups, including both software development teams and teams from other domains. Understood broadly, diversity arises from any attribute that differentiates people \cite{paper8}. These attributes can be demographic (\eg, age, gender, culture, religion, nationality), functional (\eg, role, expertise), or subjective (\eg, personality).

The impact of diversity on team outcomes remains a topic of debate, with conflicting findings in the literature. While some studies report a positive correlation between diversity and performance, others suggest a negative relationship. For instance, \cite{paper9} states that diversity can prevent grouping and lead to an inclusive workspace while also adding problems in how teammates communication. 

\subsection{Gender Diversity in Software Teams}
Said by the World Health Organization, \textit{gender} refers to ``\textit{the socially constructed characteristics of women and men – such as norms, roles and relationships of and between groups of women and men. It varies from society to society and can be changed}''
\footnote{\href{https://www.who.int/health-topics/gender\#tab=tab_1}{https://www.who.int/health-topics/gender\#tab=tab\_1}}. 
Nevertheless, there are multiple definitions, as the American Psychology Association defines \textit{gender identity} as a component of gender that describes a person’s psychological sense of their gender. Many people describe gender identity as a deeply felt, inherent sense of being a boy, a man, or male; a girl, a woman, or female; or a nonbinary gender (\eg, genderqueer, gender-nonconforming, gender-neutral, agender, gender-fluid) that may or may not correspond to a person’s sex assigned at birth, presumed gender based on sex assignment \cite{APA}.

Gender diversity has been increasingly recognized as a key factor in enhancing productivity and reducing turnover in software development teams. Studies show that gender-diverse teams tend to exhibit higher productivity, with partial positive effects on reducing employee turnover \cite{gendertenure}. This positive influence is linked to diverse perspectives in problem-solving and communication, which contribute to more dynamic and effective team interactions. The impact of gender diversity on employee performance is further influenced by several demographic and social factors, including the gender, age, marital status, education level, position, and work experience of employees. Research demonstrates a positive correlation between gender diversity and individual employee performance within organizations and statistical analyses reveal that gender diversity can be a predictive factor for employee performance, as diverse teams foster harmonious relationships and positive attitudes \cite{performance}. This, in turn, enhances overall team performance and adaptability to evolving market demands.

Another way to analyze the women's behaviour in teams is through studying its effect in \textit{community smells}, which are organizational patterns associated with problematic team dynamics and communication structures \cite{codesmells}. The top four more popular are: \textit{Organizational silos}, when subgroups within a team are isolated, leading to fragmented knowledge and collaboration breakdowns; \textit{Black clouds}, which describe persistent negative attitudes or dissatisfaction within a team, which can spread and impact morale; the \textit{Lone wolf} phenomenon, that refers to individuals working in isolation from the rest of the team, limiting collaborative potential; and cohesion and \textit{Radio silence}, that indicates a lack of communication, often resulting in misalignment and reduced efficiency in project execution. In terms of organizational culture, gender-diverse teams report fewer community smells compared to non-diverse teams \cite{communitysmells}. Specifically, gender diversity and women’s participation are influential in mitigating issues such as \textit{Black clouds} and \textit{Radio silence}. However, the effect of gender diversity on other patterns, such as \textit{Organizational silos} and \textit{Lone wolves} appears to be less pronounced. 

Nonetheless, women do not appear much in software teams; maybe because there are too few of them working in the field or maybe because of an inherent bias. The 2022 Stack Overflow Developer Survey \cite{stackoverflow} showed that only 5.17\% of worldwide developers (from 180 countries) were women, while the 2023 data from Eurostat \footnote{\href{https://ec.europa.eu/eurostat/statistics-explained/index.php?title=ICT_education_-_a_statistical_overview}{https://ec.europa.eu/eurostat/statistics-explained/index.php?title=ICT\_education\_-\_a\_statistical\_ overview}} revealed that just 17.4\% ICT employees were women. Imtiaz \etal \cite{biasGithub} conducted an investigation to examine four effects on gender bias; \textit{Tightrope}, when women behave in a restricted manner to avoid backlash, \textit{Maternal Wall}, when women who are mothers have their competence and commitment questioned, \textit{Prove-It-Again}, when women must provide more evidence than men to prove competence and \textit{Tug of War}, when women are discouraging to other women. These four phenomena were first described in Williams and Dempsey's work \cite{williamdempsey}. The investigation, performed with GitHub data, demonstrated that, while women are equally competent, they often limit their contributions to fewer projects and organizations compared to their male counterparts. Women also tend to demonstrate a more restrained approach in their professional interactions. Contrary to some expectations, the research has shown no significant differences between men and women in avoiding the posting of personal content, such as images with children, on professional platforms.

\textit{Table} \ref{tab:related-work} shows a summary of the aforementioned related work. While gender diversity in software projects has been widely studied, most research is focused on larger and established platforms like GitHub \cite{gendertenure} \cite{biasGithub} or relied on survey-based approaches \cite{google} \cite{roads}. This leaves a gap in understanding how gender diversity manifests itself in newer communities, such as ML engineers \cite{10243109} in HF. This offers an opportunity to explore gender dynamics in a new area, motivating us to investigate this understudied area.

\begin{table}[!h]
\centering
\begingroup
\footnotesize 
\begin{threeparttable}
\caption{Comparison of related work.}
\label{tab:related-work}
\begin{tabular}{p{2cm}p{2.8cm}p{5.7cm}p{3cm}} 
\toprule
  \textbf{Study} & \textbf{Focus} & \textbf{Key Findings} & \textbf{Research Strategy}\\ 
  \midrule
  \cite{paper9} & Impact of diversity on communication in teams & Diversity can prevent grouping and create an inclusive workspace but may lead to communication challenges & Field strategy \\ 
  \midrule
  \cite{gendertenure}& Gender diversity in software teams & Gender-diverse teams have higher productivity and lower turnover due to better problem-solving and communication & Data strategy \\ 
  \midrule
  \cite{performance} & Gender diversity and individual performance & Gender diversity positively correlates with individual performance, fostering better relationships and attitudes & Data strategy \\ 
  \midrule
  \cite{communitysmells} & Impact of gender diversity on organizational culture & Gender-diverse teams report fewer community smells, especially black clouds and radio silence & Data strategy \\ 
  \midrule
  \cite{biasGithub} & Gender bias in software teams & Women tend to limit their contributions and restrict professional interactions due to bias & Data strategy \\ 
  \midrule
  \cite{google} & Gender diversity in software teams & Researchers should also explore less visible challenges affecting diversity & Respondent strategy \\ 
  \midrule
  \cite{roads} & Gender diversity in software teams & Improving diversity in software development requires meaningful, sustainable solutions that balance business needs and inclusivity & Respondent strategy \\ 
  \midrule
  This study & Gender diversity in HF & Analyzes the effects of gender diversity on HF organizations and models & Data strategy \\ 
  \bottomrule
\end{tabular}
\begin{tablenotes}
    \footnotesize
    \item The last column shows the research strategy based on \cite{strategies}.
\end{tablenotes}
\end{threeparttable}
\endgroup
\end{table}

\section{Research Methodology}
\subsection{Goal and Research Questions}
Following the Goal Question Metric (GQM) template \cite{caldiera1994goal}, our goal is to \textbf{\textit{analyze} gender diversity \textit{for the purpose of} measuring it  \textit{with respect to} ML model commits \textit{from the point of view of} ML engineers \textit{in the context of} HF organizations of the HF Hub}. 
 
To gain insights into organizational structures and ML model development processes in HF, we defined four Research Questions (RQ). \textbf{RQ1} focuses on understanding the characteristics of HF organizations in terms of gender diversity to uncover patterns that define their composition. 

\textbf{\textit{RQ1: How are HF organizations built in terms of gender diversity?}}

\textbf{RQ2} and \textbf{RQ3} shift the focus to ML model commits and the contributions made by users to these projects. In this context, we aim to investigate how gender diversity within organizations relates to the number of downloads and success of the models they develop, as well as the behavioral patterns of individual contributors.

\textbf{\textit{RQ2: Does the popularity of an ML model have a relation with the gender diversity in the organization developing it?}}

\textbf{\textit{RQ3: How does gender relate to the volume and contribution of commits in HF models?}}

Building on the analysis of organizations and ML model contributions, \textbf{RQ4} compares gender diversity at the organizational level with that of individual contributors. We aim to see if gender patterns observed in organizations are also reflected in individual contributions.

\textbf{\textit{RQ4: How does gender distribution in the top individual HF contributors compare to that in leading organizations?}}

\subsection{Repository Mining Process}
\begin{figure*}[!htb]
    \centering
  \includegraphics[width=\textwidth]{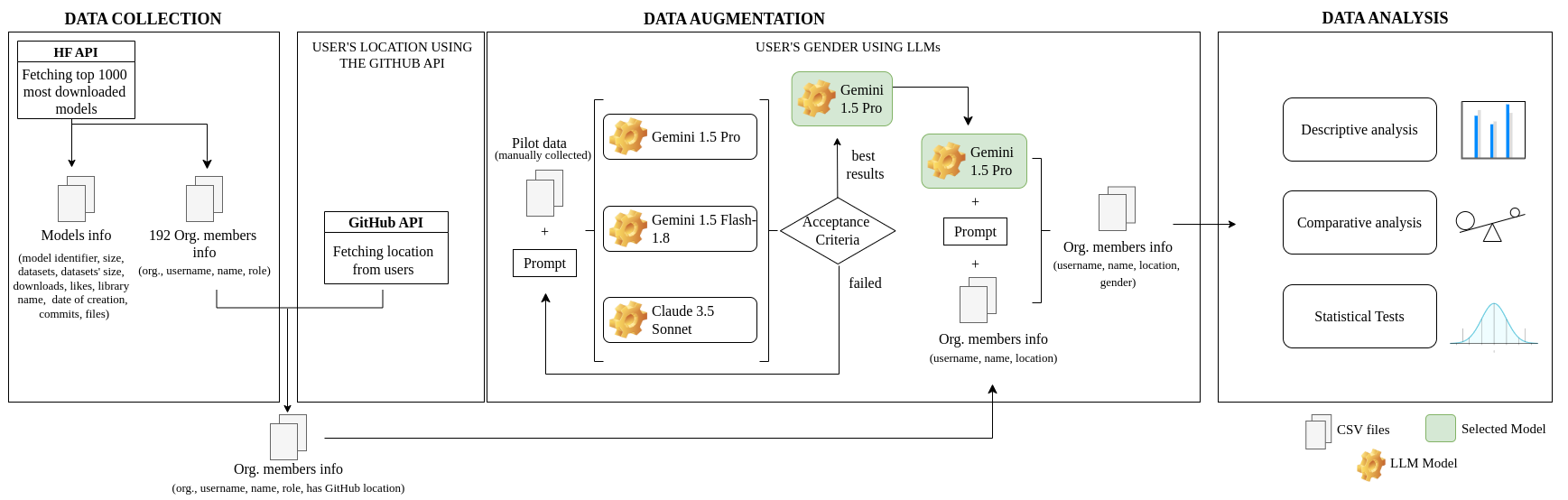} 
    \caption{Repository Mining Study: Data Collection, Augmentation, and Analysis.}
    \label{fig:diagrama}
\end{figure*}

As illustrated in \textit{Figure} \ref{fig:diagrama}, this repository mining study is divided into three fundamental parts: data collection from the HF API, the use of GitHub API and Large Language Models (LLMs) to augment that data, and a final analysis stage. 

\subsubsection{Data Collection}

The data collection process began by using the HF API \footnote{\href{https://huggingface.co/docs/huggingface_hub/package_reference/hf_api\#huggingface_hub.HfApi}{https://huggingface.co/docs/huggingface\_hub/package\_reference/hf\_api\#huggingface\_hub.HfApi}}, following methods from \cite{10.1145/3643991.3644898}, to retrieve information about users, including their affiliations with organizations and details about the models they manage. We retrieved the \textbf{top 1,000 most downloaded models}, associated with \textbf{363 different developers}, including individual users who independently published models and organizations. Since our focus was on studying organizations, we filtered the data to retain only the models associated with them, leaving us with \textbf{192 organizations} and their most downloaded models. For these organizations, we collected the necessary data. On the one hand, we obtained the data from the \textbf{models}: \textit{model identifier, size, datasets, datasets' size, downloads, likes, library name, date of creation, commits} and \textit{files}. However, to conduct our experiment we only used \textit{model identifier, downloads, likes} and \textit{commits}. Each \textit{commits} data comprised a JSON with all ML model commits registered. Each commit is composed of these different fields: \textit{commit identifier, authors, date of creation, title, message, formatted title} and \textit{formatted message}. On the other hand, we saved the information about all \textbf{users} from those 192 organizations. That resulted in a total of 14,988 users from who we got \textit{organization, username, name} and \textit{role}.

At the same time, manually, we got the \textbf{top 100 individual contributors} from December 2024. About these users, we retrieved their \textit{username} and \textit{name} and obtained their \textit{gender} by manually looking at their profile images or either of their socials.
\subsubsection{Data Augmentation}\label{llm}

\paragraph{User's location using the GitHub API}
Determining a user's gender directly from the HF API was not possible, as the platform does not include a specific field for users to disclose this information. However, the profile configuration page allows users to link external accounts, such as GitHub, X (formerly Twitter), LinkedIn, and Bluesky. While the API does not provide access to this data, it is possible to manually infer a user’s gender by examining their profile photo on HF or by exploring these linked accounts.

Nevertheless, searching for the gender one by one for almost 15,000 people would mean a huge amount of time. In order to try to get the gender information we used the GitHub API. Because some users had their GitHub account linked, we used it to save their \textit{location}. Again, obtaining gender information was not accomplished. Many are trying to but it seems like the pronouns are highly protected and, apart from doing it manually, it is not viable to know them. So, finally, the organization's members data we collected were \textit{organization, username, name, role, has github} and \textit{location}.

\paragraph{User's gender using LLMs}
Because the experiment's goal is the study of gender diversity, it is essential to know that about the users that compose an organization. To accomplish that, we relied on LLMs and followed the PRIMES framework \cite{primes}. We focused on LLMs that provide an accessible API and offer some form of free credit for testing. Following a careful evaluation of available models, we selected Anthropic's \textit{Claude 3.5 Sonnet}, which provides a \$5 free credit, and Google's \textit{Gemini 1.5 Flash} and \textit{Gemini 1.5 Pro}, which offer \$300 trial. 
To ensure clarity and reduce potential ambiguities arising from international naming conventions, we followed the recommendation by \cite{howtoask} to provide additional context beyond just names. We conducted a pilot using a manually collected \textbf{dataset of 360 randomly selected user samples} to validate the use of the LLM, and through several prompt refinements, we derived the following prompt to complete the dataset:

\begin{tcolorbox}[colback=gray!5!white, colframe=gray!75!black, title=Final Prompt, rounded corners]
Given the name [name] and, if not none, the location [location], tell me the gender of the username.

\begin{itemize}
    \item The input is a name and maybe a location too.
    \item The output must be `Male', `Female', or `Unknown'. Without any explanations or reasons. Just the predicted gender.
    \item Examples: 
    \begin{itemize}
        \item Christopher Pirillo: Male. Vera AxelRod: Female
    \end{itemize}
\end{itemize}
\end{tcolorbox}

We chose to introduce a third category, \textit{unknown}, to prevent the LLM from being forced to classify individuals as either male or female. This approach was intended to mitigate the risk of the model providing correct answers purely by chance. This third category allows for more nuanced interpretations of results and reduces the potential for biased outcomes.
To evaluate the accuracy, we used \textbf{Cohen’s Kappa} \cite{cohen1960coefficient}, a statistical measure of agreement between two dependent categorical samples. An important consideration when interpreting the kappa is that the pilot data was built using additional information extracted from other platforms (such as HF, X and LinkedIn profile photos, published posts, Google Scholar pages and other socials appearances) while the models could only rely on name and location. We compared our manually collected data with the LLM gender classification to evaluate how well the model did.
As shown in \textit{Table} \ref{tab:llms_pilot}, the \textit{Claude 3.5 Sonnet} requires more than twice the execution time of \textit{Gemini’s Flash} and nearly four times that of the \textit{Pro}. Additionally, the free credit provided by \textit{Anthropic} would be insufficient to process our entire dataset. Despite the comparable accuracy levels across all models, execution time and economic implications led us to select \textbf{Gemini 1.5 Pro}. 

With the LLM, we augmented our dataset, which comprised each user's \textit{organization, username, name, role, has github, location} and \textit{gender}. 


\begin{table}[h]
\centering
\small
\caption{Comparison of LLMs in the pilot test.}
\label{tab:llms_pilot}
\begin{tabular}{cccc} 
  \toprule
  \textbf{LLM} & \textbf{Execution Time} & \textbf{Economic Cost} & \textbf{Cohen's Kappa} \\ 
  \midrule
  Claude 3.5 Sonnet & 12' & $\approx$ \$0.64 & Moderate Agreement \\ 
  \midrule
  Gemini 1.5 Flash & 5'40" & $\approx$ \$0.50 & Moderate Agreement \\ 
  \midrule
  Gemini 1.5 Pro & 3'46 & $\approx$ \$0.50 & Moderate Agreement \\ 
  \bottomrule
\end{tabular}
\begin{tablenotes}
    \footnotesize
    \item Explicit Cohen's Kappa values can be found in the Replication package \ref{ArtAv}.
\end{tablenotes}
\end{table}

\subsubsection{Data Analysis}
We used \textbf{descriptive analysis} to explore patterns of diversity and contributions within organizations. This method quantified important aspects such as gender distribution and commit characteristics, offering an overview of trends. Additionally, \textbf{comparative analysis} contrasted gender distributions and contributions between organizations and individual developers, providing deeper insights into diversity dynamics on the platform. Finally, we conducted \textbf{statistical tests} for each RQ, with a significance level of $\alpha = 0.05$ for all analyses. For \textit{RQ1}, we performed a $\chi^2$ test to determine whether there is an association between organizations and gender distribution across the dataset. To address \textit{RQ2}, we investigated the relationship between the proportion of women in an organization and the downloads of their most popular model, applying Spearman’s correlation test due to the non-normal distribution of the data. For \textit{RQ3}, we used the Kruskal-Wallis test to examine differences in the distribution of commits between men, women, and bots, as the normality assumption was not met. Lastly, for \textit{RQ4}, we conducted a $\chi^2$ test to explore the relationship between gender distribution in organizations and among individual contributors. The statistical tests and the null hypotheses are summarized in \textit{Table} \ref{tab:rq_hypotheses}.

\begin{table}[h!]
\centering
\small
\caption{Summary of RQs, Statistical Tests, and Null Hypothesis (\(H_0\)).} 
\label{tab:rq_hypotheses}
\begin{tabular}{ccc}
\toprule
\textbf{RQ} & \textbf{Test}& \textbf{Null Hypothesis (\(H_0\))}\\ \midrule
RQ1 & \(\chi^2\) test & \makecell{No association between organization and gender.} \\ \midrule
RQ2 & Spearman's correlation & \makecell{No correlation between women's proportion and\\ ML model downloads.}\\ \midrule
RQ3& Kruskal-Wallis test & \makecell{Commit distribution is equal across groups.}\\ \midrule
RQ4& \(\chi^2\) test & \makecell{Organization and contributor gender distributions \\are the same.}\\ \bottomrule
\end{tabular}
\end{table}

\section{Results} \label{results}
\subsection{Gender diversity in HF organizations (RQ1)}

As illustrated in \textit{Table} \ref{tab:users_gender_org}, women account for 12.06\% of the total users across the 192 HF organizations analyzed \footnote{\href{https://github.com/GAISSA-UPC/HF_diversity/tree/main/HF_diversity/HF_diversity/organisations/ complete_organisations}{https://github.com/GAISSA-UPC/HF\_diversity/tree/main/HF\_diversity/HF\_diversity/organisations/ complete\_organisations}} when including those with unknown gender and 17.21\% when considering only users with an assigned gender. A closer examination reveals that the proportion of women in the studied organizations ranges from 0\% to 50\%, with no organization having a female majority. Notably, 88 organizations have no women on their teams, and 183 have less than 30\% female representation. Among the few organizations where women constitute half of the team, all consist of four members or fewer, underscoring the limited female presence across these organizations.

\begin{table}[h]
\centering
\small
\caption{Users by gender in the 192 analyzed organizations}
\label{tab:users_gender_org}
\begin{tabular}{cccc}
\toprule
   \textbf{Gender} & \textbf{Count} & \textbf{Proportion (\%)} & \textbf{Mean Users}\\
  \midrule
  Men & 8,696 & 58.02\% & 45.29\\ 
  Women & 1,808 & 12.06\% & 9.42\\ 
  Unknown & 4,484 & 29.92\% & 23.35\\ 
  \midrule
  Total & 14,988 & 100\% & 78.06\\ 
  \bottomrule
\end{tabular}
\end{table}

\noindent
\colorbox{orange!20}{\parbox{\linewidth}{\textbf{Finding 1.} There is a significant gender imbalance within HF organizations, with the number of women remaining low, even when considering only users with known gender.}}

The $\chi^2$ test yielded a p-value of $3.28 \times 10^{-57}$, which is far smaller than the significance level. Consequently, we rejected the null hypothesis, concluding that the proportions of males, females, and unknown gender differ significantly across organizations. The result is consistent with the gender imbalance observed across the different HF organizations, as well as with the diverse team sizes and highly varied gender distributions.

\subsection{Relationship between ML model popularity and gender diversity  in HF Organizations (RQ2)}

\begin{figure}[h]
  \begin{center}
  \includegraphics[width=0.5\textwidth]{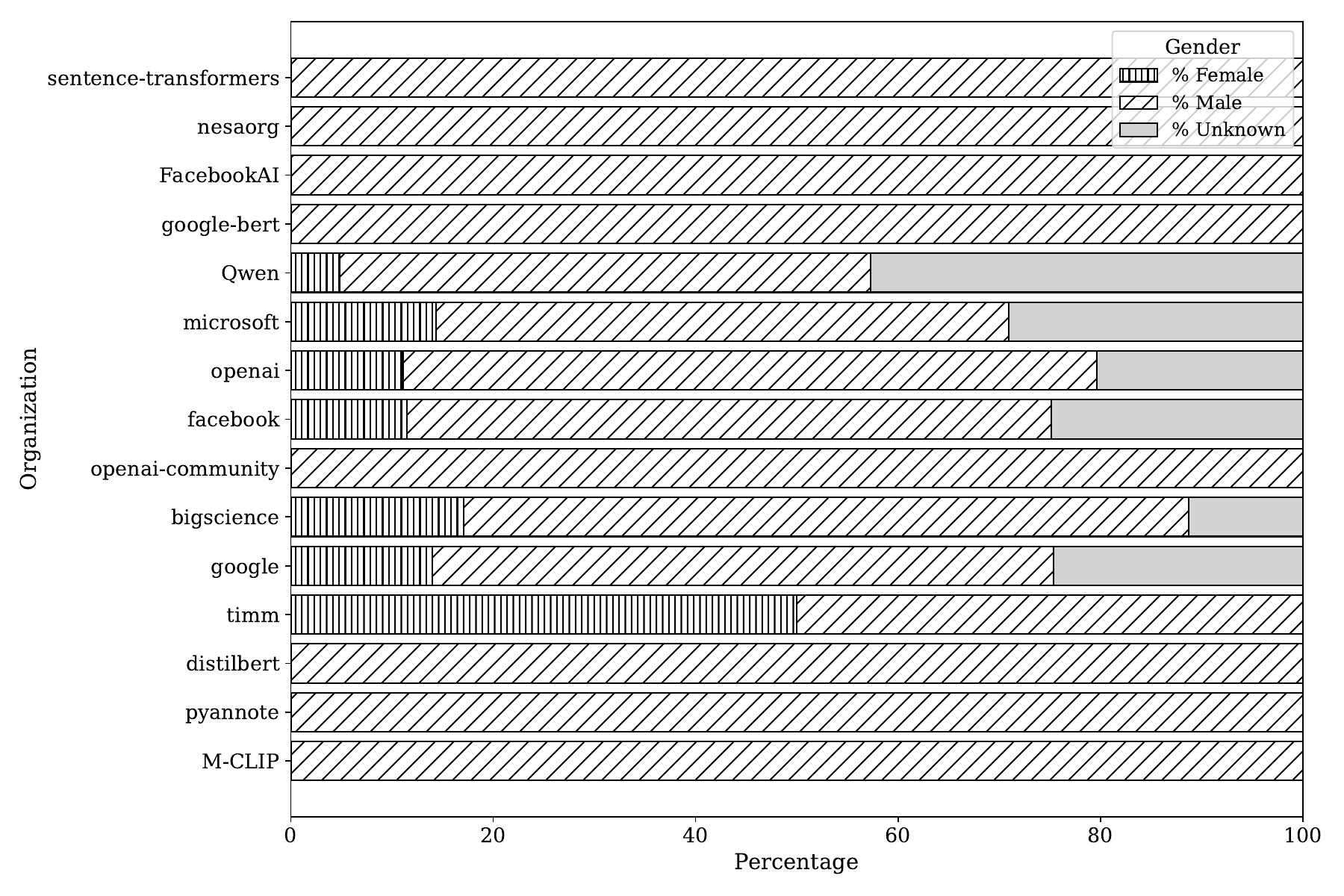} 
    \caption{Normalized Gender Distribution of the Top 15 Organizations.}
    \label{fig:genderdistr}
    \end{center}
\end{figure}

We selected the \textbf{most downloaded model} from each organization to explore whether there is a relationship between ML model popularity and gender diversity. We then examined the behavior of the top 15 organizations - organizations that developed the 15 models with the highest download counts - including: \textit{sentence-transformers, nesaorg, facebookAI, google-bert, Qwen, microsoft, openai, facebook, openai-community, bigscience, google, timm, distilbert, pyannote} and \textit{M-CLIP}. \textit{Figure} \ref{fig:genderdistr} shows no evident correlation between team diversity and the number of downloads. In fact, the top 3 organizations, with the highest download counts, have no women on their teams, suggesting a potential inverse relationship between gender diversity and popularity. Among the top 15 organizations, more than half are composed entirely of male members, highlighting the absence of gender representation in the most successful teams. These observations challenge the assumption that diversity is positively associated with organizational success, at least in terms of download metrics. However, it is important to note that the popularity of an ML model does not necessarily equate to the popularity of its organization, as models are often developed by contributors outside of the organization. Thus, it would not be accurate to assume that the gender distribution within a team aligns with that of the ML model authors.

\noindent
\colorbox{orange!20}{\parbox{\linewidth}{\textbf{Finding 2.1.} The most downloaded models do not exhibit a direct relationship with organizational diversity, as the top organizations often have no female team members.}}

Next, we analyzed \textbf{all 192 models} rather than just the top ones per organization, and obtained Spearman’s correlation coefficient of -0.03 and a p-value of 0.7. The correlation value indicates a very weak negative relationship between the proportion of women and the number of downloads. Since the correlation is close to zero, it suggests that there is no clear relationship between the two variables. The negative direction implies that, on average, as one variable increases, the other decreases slightly, but this trend is negligible due to the small correlation value. Furthermore, the p-value of 0.7 is much greater than the significance threshold of 0.05, providing no evidence to reject the null hypothesis ($H_0$). Therefore, we conclude that there is no significant correlation between the proportion of women and the number of downloads.

\noindent
\colorbox{orange!20}{\parbox{\linewidth}{\textbf{Finding 2.2.} No significant correlation exists between the proportion of women and ML model downloads across all models.}}

\subsection{Commits data and gender diversity (RQ3)}

We further analyzed the commit distribution of models developed by the 192 organizations selected for this study. However, before diving into the analysis, several key observations were made.

Firstly, we observed that the commit distribution does not directly correspond to the distribution of organizations. Due to the open-source nature of the platform, individuals can contribute to models without being affiliated with a particular organization or owning the model. As a result, some users make contributions to over 30 models, despite not being members of any team. Additionally, we encountered what we refer to as \textbf{Bots}, which contribute a significant number of commits. However, these commits are not critical to ML model development, as they mainly focus on file-related tasks. Two notable bots, \textit{system} and \textit{SFconvertbot}, rank as the top two committers. The \textit{system} bot is responsible for managing ML model files such as \textit{.txt, .json,} and \textit{.bin}, as well as tasks such as updating \textit{README.md} or making the initial commit. In contrast, the \textit{SFconvertbot} specializes in a single type of commit: \textit{“Adding ‘safetensors’ variant of this model”}.

In our analysis of commit data, we recorded 5,274 commits across 192 distinct models (one per organization). The data necessary to complete the final component of this experiment was collected manually due to the nature of authors’ independent commitments to models. Consequently, when combining data from the organizations with commit data, many authors could not be identified because the LLM did not process those users.

\begin{figure}[h]
\begin{center}
    \begin{minipage}{0.49\textwidth}
        \centering
        \includegraphics[width=\textwidth]{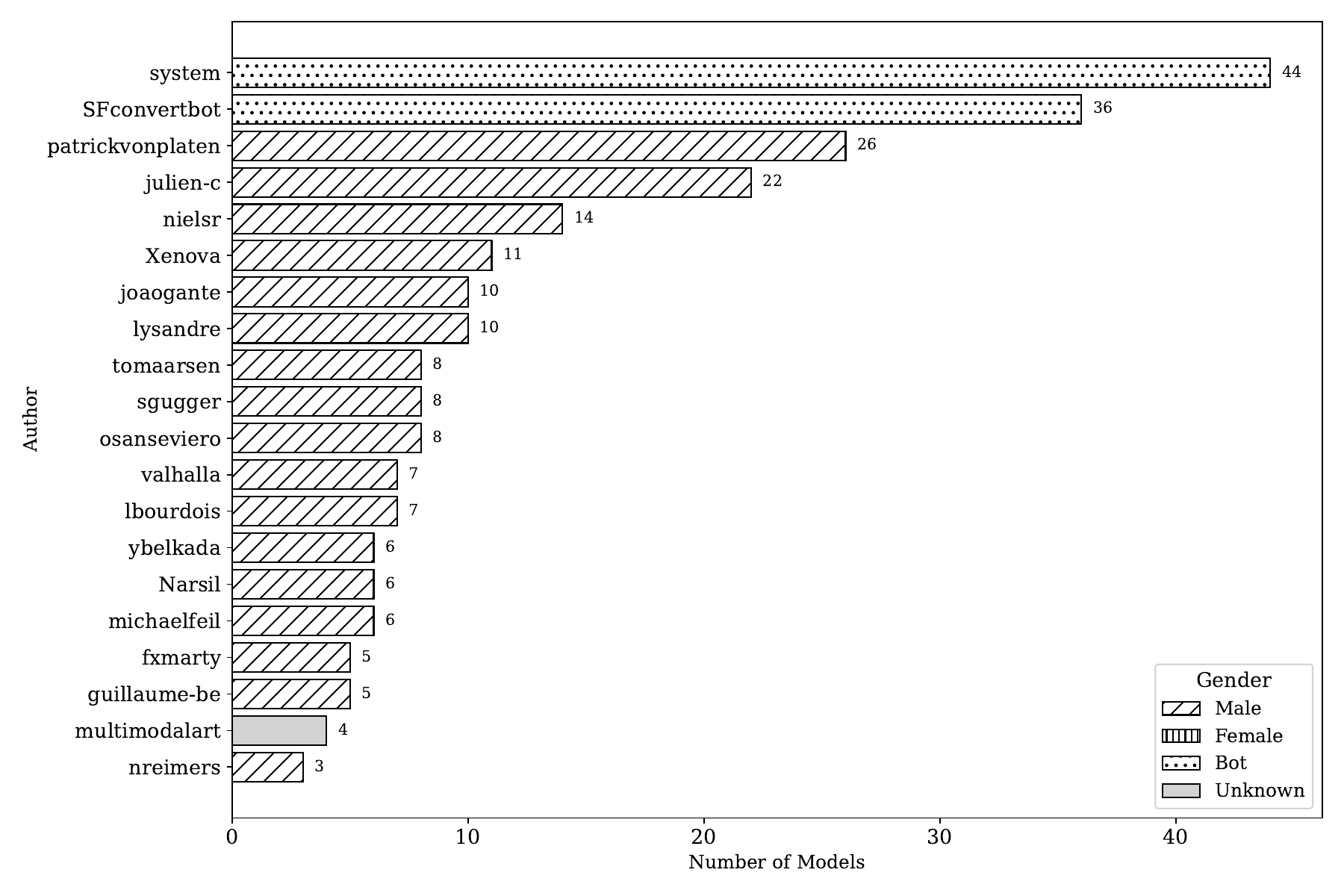}
        \caption{Top 20 Authors by Number of Models they Committed in.}
        \label{fig:topnummodels}
    \end{minipage}
    \hfill
    \begin{minipage}{0.49\textwidth}
        \centering
        \includegraphics[width=\textwidth]{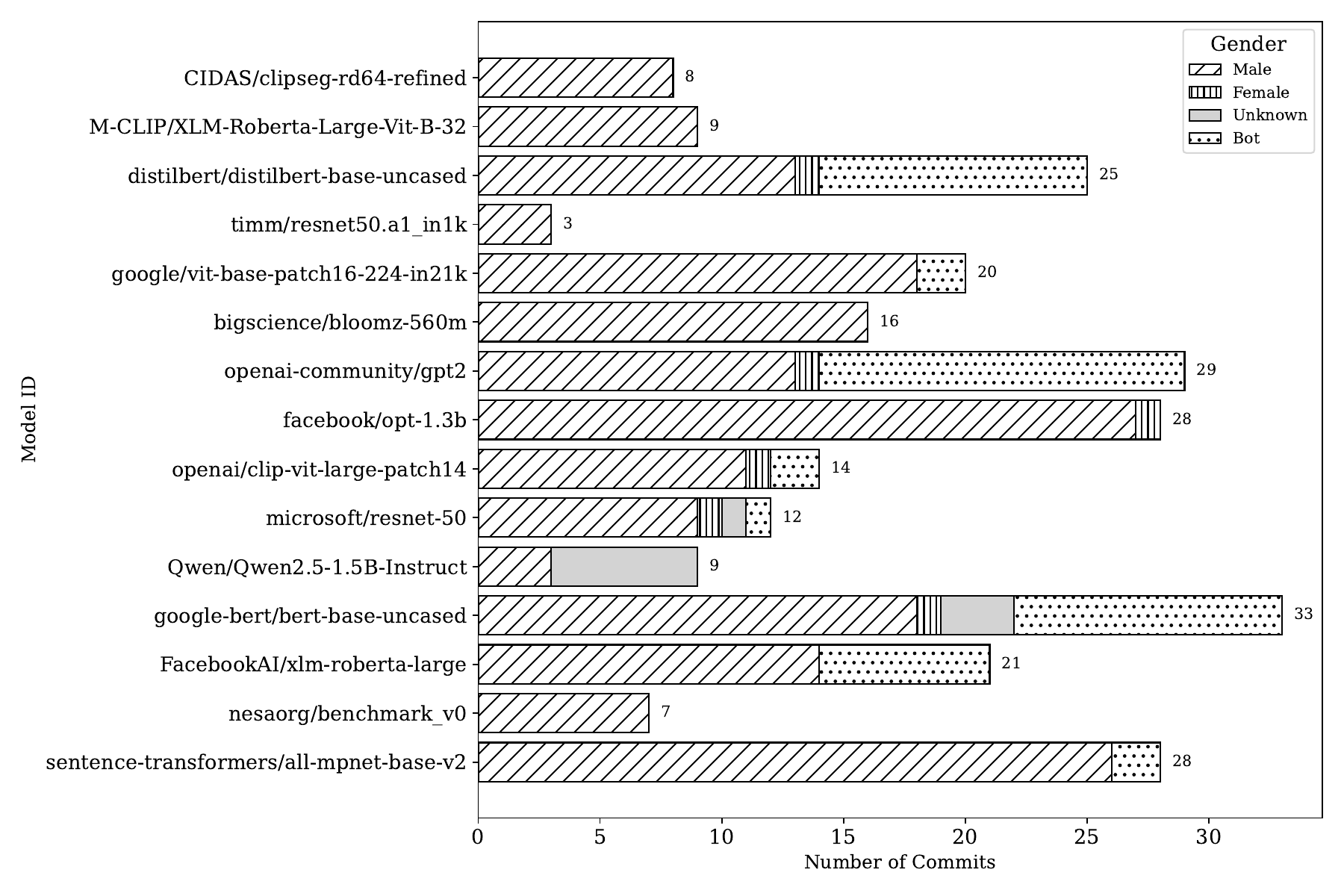}
        \caption{Number of Commits by Gender for the Top 15 Most Downloaded Models.}
        \label{fig:top15_model}
    \end{minipage}
\end{center}
\end{figure}

Initially, we analyzed the gender distribution among authors who had committed the most, based on the number of models they contributed to. As observed in \textit{Figure} \ref{fig:topnummodels}, there were no female authors among the top committers, with only five female authors contributing to a total of two models. Further examination of the authors with the highest total number of commits revealed a similar lack of female representation. Of the 5,274 commits collected, 83.35\% were done by men, 6.48\% by users of unknown gender, 5.84\% by bots, and the remaining 4.3\% were committed by female authors.

Recovering the previous analysis - \textit{Figure} \ref{fig:genderdistr} - we explored how are the commits \footnote{pyannote's ML model was not authorized to download, that's why CIDAS appears} of their respective models (\textit{Figure} \ref{fig:top15_model}). As stated before, both gender arrangements do differ. When comparing both perspectives - one from the organization point of view and the other from their corresponding ML model- some differences can be found. The first one and obvious, bots are not part of any organization yet they commit in majority. Furthermore, in some cases, there are women committers even though that team is male-composed. That would be the case of \textit{google-bert}'s \textit{bert-base-uncased}, \textit{openai-community}'s \textit{gpt2} or \textit{distilbert}'s \textit{distilbert-base-uncased}. The contrary phenomenon appears too, as the models \textit{Qwen2.5-1.5B-Instruct}, \textit{bloomz-560m}, \textit{vit-base-patch16-224-in21k} and \textit{resnet50.a1-in1k} show no commits from female authors besides having women as members of the developing organization. 
 
\noindent
\colorbox{orange!20}{\parbox{\linewidth}{\textbf{Finding 3.1.} The gender distribution of committers does not always align with the gender composition of the developing organizations. 

\textbf{Finding 3.2.} The most committers are men, with bots making more commits than women.}}
 
Despite the significant disparity in the quantity of contributions between men and women, we examined whether any differences could be observed in the nature of the commits rather than just their volume. By analyzing the commits' title, we categorized them into two main groups: commits related to ML model changes and those related to documentation. ML model updates such as fixing or adding variables, tokenizers, or weights were classified under the first group, while tasks such as creating and updating a README.md or a model card would be included in the second. 



From a broader perspective, after conducting the Kruskal-Wallis test, we obtained a p-value of 0.008, which is smaller than the significance level of 0.05. Consequently, we rejected the null hypothesis and concluded that there are significant differences in the commit patterns across the different groups. These results are consistent with our earlier findings, where we observed that the number of commits made by men is significantly greater than those made by women or bots.

\subsection{Comparison with top users outside HF organizations (RQ4)}
To examine how the gender distribution in top organizations compares to that of individual contributors, we analyzed the \textit{Top Contributors to Follow} \cite{huggingfaceContributorsFollow}. This leaderboard, which ranks users based on metrics such as ML model downloads, dataset downloads, space likes, and profile followers, provides insight into individual developers. By focusing on ML model downloads, we applied the same approach used for organizations to study the distribution among top contributors. As shown in \textit{Table}~\ref{tab:top_contr_org}, the gender distribution among the top contributors closely resembles that of the leading organizations. While some differences exist, both groups follow a similar pattern in terms of representation across genders.

\begin{table}[h]
\centering
\small
\caption{Gender Distribution in Top 100 Contributors and Top 192 Organizations.}
\label{tab:top_contr_org}
\begin{tabular}{cccc}
\toprule
   \textbf{Category} & \textbf{Men}& \textbf{Women}& \textbf{Unknown}\\
  \midrule
  Top 100 Contributors & 60.0\% & 9.0\% & 31.0\%\\ 
  Top 192 Organizations & 58.0\% & 12.1\% & 29.9\%\\ 
  \bottomrule
\end{tabular}
\end{table}

\noindent
\colorbox{orange!20}{\parbox{\linewidth}{\textbf{Finding 4.} Gender distribution is similar across organizations and individual developers.}}

Lastly, the Chi-squared test yielded a p-value of 0.644, leading us to not reject the null hypothesis, indicating no significant differences in gender distributions. These results align with the earlier findings, as shown in \textit{Table} \ref{tab:top_contr_org}.

\section{Discussion}
Our analysis shows that women are underrepresented in HF organizations. Contrary to what Vasilescu \etal stated in \cite{gendertenure}, we found that an ML model success is not directly related to the gender diversity in it. However, we can agree with Imtiaz \etal, whose findings in \cite{biasGithub} exhibit that ``women focus their work in fewer projects''. This observation does not necessarily reflect a barrier to participation in open-source coding and development. Open-source platforms inherently allow anyone to contribute regardless of organizational affiliation. Our study reveals that model popularity does not consistently correlate with organizational diversity, as many leading organizations lack female representation. Furthermore, the gender makeup of contributors often diverges from the organization’s internal gender composition, indicating that external contributions may play a significant role in shaping widely adopted models. This means that while women may not be prominent in formal organizations, their contributions to the open-source ecosystem can still be significant and impactful. 

However, understanding the true extent of this impact requires more than just counting commits. It is crucial to note that the quantity of commits cannot be assumed to be proportional to the quality or significance of the work performed. As we only have access to commit \textit{titles} and \textit{messages}, we cannot assure the actual impact of each contribution. Therefore, it would be unfair to infer that men’s work is more relevant or substantial than women’s based solely on commit counts. It would be valuable to develop measures that assess the relevance of a commit — such as its influence on project development, the number of dependencies affected, or the extent of community engagement it generates. Such an approach could provide deeper insights into the meaningful contributions of women in open source beyond sheer participation rates.

Interestingly, the patterns observed in HF organizations appear to mirror those seen in the broader context of software development. A similar lack of gender diversity is evident when examining the most recognized individual contributors on the platform. What is more, this is a general tendency in the SE field. Survey-method work done by Kohl and Prikladnicki in \cite{gender-diversity-book} \cite{book} found that respondents answered ``it is common to see only one woman in teams or just a few''. 

The nature of open-source coding has posed challenges in analyzing the relationship between team composition and ML model commits. The decentralized and collaborative nature of open-source projects has made it difficult to reliably relate contributions to specific individuals or teams. Consequently, our findings may lack the precision needed to draw definitive conclusions, and some of our results could be considered insignificant or unreliable due to these limitations.

\section{Threats to Validity}
In this section, we address potential threats to the study’s validity, aiming to clarify the constraints and biases that could affect the interpretation of the findings.

\textbf{Internal Validity}: Since HF does not provide gender data, we relied on an LLM using names and locations to infer gender. To mitigate biases, we conducted a pilot and measured agreement using Cohen’s Kappa. The results, however, indicate that this method cannot reliably replace human judgment, highlighting the need for human involvement to ensure accuracy in this context. Nevertheless, even though we classify some users as having an \textit{unknown} gender due to insufficient information, if these users were indeed women, the proportions would still remain unbalanced. Additionally, analyzing contributions made by the same individual across multiple models has proven challenging. To address this, we aggregated all contributions by a user, regardless of their affiliation, ensuring a comprehensive perspective on their involvement across the platform.

\textbf{External Validity}: Our findings, based on HF organizations, may not fully generalize to all open-source ML communities or software ecosystems. However, by focusing on a prominent ML repository with a diverse range of organizations, our findings aim to be significant for the community. We analyzed the top 1,000 most downloaded models to assess whether model popularity is associated with gender diversity among contributors. However, the findings may not extend to less popular or niche models.

\textbf{Construct Validity}: Our study focuses on women’s representation in HF organizations, using inferred gender data rather than self-reported identities, which may not fully capture non-binary contributors or broader diversity factors. As gender is not reported, we focus on women’s roles, recognizing the existence of non-binary identities.

\textbf{Conclusion Validity}: Due to the absence of a gender field in HF, we inferred gender based on names and locations, considering regional variations. Despite potential variability, the study’s conclusions are expected to remain reliable.

\section{Conclusion and Future Work}
Our study has revealed a significant gender imbalance within HF organizations, with women being notably underrepresented. However, no direct relationship was found between gender diversity in organizations and the popularity of the models they develop. Additionally, gender diversity was not present in the participation of committers, with men being the primary contributors. Despite this, no differences were observed in the quality or content of commits between men and women, indicating that women's contributions are not limited to specific tasks or roles. Their commits span a wide range of development activities, including core functionality, bug fixes, documentation, and feature enhancements. It is important to note that this analysis is based on commit titles and messages, which serve as the primary data for this study. Similar gender trends were observed among individual contributors, reflecting those in organizations. These findings highlight the need for greater gender inclusion in open-source communities, while also showing that gender does not influence the quality of contributions.

Future research could build upon this study by analyzing a broader range of models and organizations within HF. This would provide a wider and more representative view of gender diversity and contribution patterns. Moreover, integrating user feedback through surveys could shed light on the factors influencing contributors’ decisions to engage in ML model development, helping to identify potential barriers or motivators for participation. Lastly, future studies could concentrate on evaluating the relevance and impact of commits by moving beyond surface-level attributes to gain deeper insights into their true contribution to ML model development and overall success. By doing so, we could be able to study different types of commits and committers, as in \cite{profiles}.
\section{Artifact availability} \label{ArtAv}
If wanted to reproduce the study, used codes and generated datasets can be found in this GitHub repository: \href{https://github.com/GAISSA-UPC/HF\_diversity}{https://github.com/GAISSA-UPC/HF\_diversity}.

\section*{Acknowledgment}
This work has been funded by the Spanish research project GAISSA (TED2021-130923B-I00 by MCIN/ AEI/10.13039/501100011033).

\bibliographystyle{sbc}
\bibliography{references}

\end{document}